\documentclass[aps, prb, showpacs, preprint, preprintnumbers, groupedaddress]{revtex4-1}

\usepackage[utf8]{inputenc}
\usepackage{graphicx}
\usepackage{rotating}
\usepackage{dcolumn}

\begin{document}


\title{Anisotropy in the dielectric function of Bi$_2$Te$_3$ from first principles: From the UV-visible to the infrared range}

\author{R. Busselez$^{1}$, A. Levchuk$^{1}$,
P. Ruello$^{1}$, V. Juv\'e$^{1}$ 
and B.~Arnaud$^{1}$}

\affiliation{$^{1}$
Institut des Mol\'ecules et Mat\'eriaux du Mans, 
UMR CNRS 6283, Le Mans universit\'e,
72085 Le Mans, France, EU}




\date{\today}

\begin{abstract}
The dielectric properties of Bi$_2$Te$_3$, a layered compound crystallizing in
a rhombohedral structure,
are investigated by means of first-principles calculations at 
the random phase approximation level. A special attention is devoted
to the anisotropy in the dielectric function and to the local field
effects that strongly renormalize the optical properties in the UV-visible
range when the
electric field is polarized along the stacking axis. Furthermore,
both the Born effective charges for each atom and the zone center
phonon frequencies and eigenvectors needed to describe the dielectric response
in the infrared range are computed. Our theoretical near-normal incidence 
reflectivity spectras in both the UV-visible and infrared range are 
in fairly good agreement with the experimental spectras, provided that
the free carriers Drude contribution arising from defects is included in 
the infrared response. The anisotropic plasmon frequencies entering the
Drude model are computed within the rigid band approximation, suggesting
that a measurement of the reflectivity in the infrared range for both
polarizations might allow to infer not only the type of doping but also
the level of doping.   
\end{abstract}


\maketitle

\section{INTRODUCTION\label{intro}}
A great deal of interest has been focused on Bi$_2$Te$_3$ because
of its outstanding thermoelectric properties at room temperature
arising from both a narrow gap electronic structure and a low thermal
conductivity\cite{snyder_2008}. The more recent discovery of three-dimensional topological
insulators has renewed the interest for this compound. Indeed, soon after
Bi$_2$Te$_3$ was theoretically predicted to host topologically non trivial 
metallic surface states with a Dirac-like dispersion around 
the surface Brillouin zone center\cite{zhang_2009}, some angular-resolved photoemission spectroscopy 
(ARPES) studies revealed the existence 
of such linearly dispersing surface states\cite{chen_2009, hsieh1_2009, hsieh2_2009}.
While these extraordinary electronic properties encoded in the bulk wavefunctions 
have been thoroughly 
investigated both experimentally\cite{kohler1_1976, kohler2_1976, mohelsky_2020, hajlaoui_2012, weis_2021} 
and theoretically\cite{wang_2007, luo_2012, kioupakis_2010, yazyev_2012, aguilera_2013, nechaev_2013},
the infrared (IR) and optical responses have been studied 
more sparsely\cite{richter_1977, chapler_2014, Greenaway_1965, dubroka_2017}, notwithstanding 
the fact that applications in the technology relevant IR to visible range have
been envisionned. 
Indeed, experiments have shown that Bi$_2$Te$_3$ can support surface plasmons
in the entire visible range\cite{zhao_2015} and some ellipsometry measurements even
suggest that Bi$_2$Te$_3$ is a natural hyperbolic material in the near-IR to visible
spectral range with appealing applications like hyperresolved superlensing\cite{esslinger_2014}. 
Surprisingly, the optical properties infered from reflectivity 
spectras\cite{Greenaway_1965} or ellipsometry 
measurements\cite{dubroka_2017, zhao_2015, esslinger_2014} are rather scattered. 
Hence, it's important to compute reference optical properties at a given level of theory\cite{onida_2002} that can be directly compared to the available experimental results. Such a confrontation between theory and experiment is crucial not only to refine the level of theory needed to achieve a good description of the experimental results but also to spur new experiments guided by theory.

The paper is organized as follows. In section \ref{details}, 
we give an account of the technicalities used to perform our first-principles calculations.
In section \ref{structure}, we describe the crystallographic structure
of Bi$_2$Te$_3$ and compare our calculated lattice constants and internal coordinates
to the experimental values extracted from X-ray measurements.
In section \ref{electronic_structure}, we discuss the rather complex bandstructure
including spin-orbit coupling (SOC) computed within the local density 
approximation (LDA) and
show that the band gaps, as expected and already demonstrated in other theoretical
studies\cite{kioupakis_2010, yazyev_2012, aguilera_2013, nechaev_2013}, 
are underestimated with respect to the values infered from 
optical spectroscopy measurements.  
In section \ref{optical_properties}, we compute the optical properties within the random
phase approximation (RPA) for an electric field either parallel or perpendicular to the
stacking axis, discuss the crucial role of local field (LF) effects and
make a direct comparison with available experimental results.
In section \ref{frequencies}, we compare our calculated zone center phonon frequencies
to the frequencies extracted from both Raman and infrared (IR) spectroscopy measurements.
In section \ref{Born_Charges}, we compute the Born effective charge tensors that are 
key ingredients to compute the IR dielectric tensor.
In section \ref{dielectric_tensor}, we derive the formalism used to compute 
the IR dielectric tensor with our home made code and discuss the anisotropy in our computed 
dielectric tensor for frequencies in the THz range. In section \ref{IR_reflectivity}, 
we compute the
normal incidence reflectivity spectras for both polarizations 
and only partially
reproduce the experimental spectras because free carriers arising from
defects contribute to the IR dielectric tensor and lead to a drastic change in the
calculated spectras. Finally, in section \ref{plasmons}, we compute the anisotropic plasmon
frequencies using a rigid band approximation for different type
and level of doping. A direct comparison with the fitted plasmon frequencies 
used in part \ref{IR_reflectivity} to reproduce the experimental 
spectras seems to indicate that
it might be possible to infer not only the type of doping but also the level of
doping from our calculations.

\section{COMPUTATIONAL DETAILS}\label{details}
All the calculations are done within the framework of the local 
density approximation (LDA) for the exchange-correlation functional 
to density functional theory (DFT) using the ABINIT code\cite{gonze_2009, gonze_2016}. 
Relativistic separable dual-space Gaussian pseudo-potentials\cite{hartwigsen_1998} 
are used with Bi ($6s^2 6 p^3$) and Te ($5s^2 5 p^4$) 
levels treated as valence states. Spin-orbit coupling is included and 
an energy cut-off of 40 Hartree in the planewave expansion of wavefunctions 
together with a $16\times 16\times 16$ kpoint grid for the Brillouin zone 
integration are used.

\section{LATTICE PARAMETERS AND INTERNAL COORDINATES}\label{structure} 

\begin{table}[!hbt]
\caption{Calculated LDA lattice parameters  and internal parameters including spin-orbit coupling (SOC) 
compared to both existing experimental\cite{francombe_1958, wiese_1960} 
and theoretical\cite{luo_2012} results.} 
\label{lattice_tab}
\begin{tabular}{lccccccc}
\hline
   &\multicolumn{3}{c}{Rhombohedral structure} &\multicolumn{2}{c}{Hexagonal structure}& \multicolumn{2}{c}{Internal parameters} \\
\hline
 & $a_0$ (\AA) & $\alpha_0$ ($^o$) & $V_0$ (\AA$^3$) & $a_{\parallel,0}$ (\AA) & $a_{\perp,0}$ (\AA) & $\nu$ (Te) & $\mu$ (Bi) \\ 
Expt.\cite{francombe_1958, wiese_1960} & 10.473 &  24.159 & 169.10    &30.487 & 4.383 & 0.2095 &  0.4001 \\
This work (0 K)  & 10.214    & 24.706 & 163.71     &29.692  & 4.370  & 0.2071    & 0.4009  \\     
LDA+SOC\cite{luo_2012}  & 10.124    & 24.806 & 160.65     &29.423  & 4.349  & 0.2063    & 0.4012  \\     
This work (300 K) & 10.473 &  24.159 & 169.10    &30.487 & 4.383 & 0.2091 &  0.4002 \\
\hline
\end{tabular}
\end{table}

\begin{figure}[!htbp]
\begin{center}
\includegraphics[angle=0, scale=0.4]
{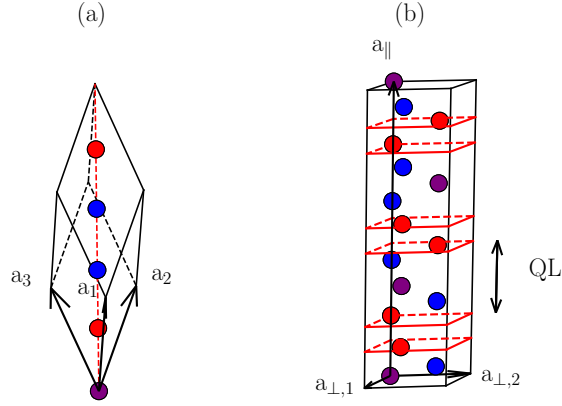}
\end{center}
\caption{
(a) Rhombohedral structure with 5 atoms per unit cell.
The two Te$_1$ atoms, the Te$_2$ atom and the two Bi atoms are respectively colored in red,
purple and blue.
(b) Hexagonal structure with 15 atoms per unit cell. The planes containing Te$_1$ atoms
are highligthed and one quintuple layer (QL) Te$_1$-Bi-Te$_2$-Bi-Te$_1$ is shown.
}
\label{structure_fig}
\end{figure}

Bi$_2$Te$_3$ crystallizes in a rhombohedral structure, also called A7 structure, 
with five atoms per unit cell. The vectors spanning the unit cell are given by
\begin{equation}
{\bf{a}}_1=\left(a \xi, -\frac{a \xi}{\sqrt{3}}, h \right) ;
{\bf{a}}_2=\left(0, \frac{2 a \xi}{\sqrt{3}}, h \right) ;
{\bf{a}}_3=\left(-a \xi, -\frac{a \xi}{\sqrt{3}}, h \right),
\end{equation}
where $\xi=\sin[\frac{\alpha}{2}]$ and $h=a\sqrt{1-\frac{4}{3}\xi^2}$. The length
of the three lattice vectors is equal to $a$ and the angle between any pair of vector 
is $\alpha$. The three Te atoms can be classified into two inequivalent types. Two of them,  
labelled as Te$_1$, are located at $\pm \nu {\bf{c}}$ while the last
Te atom, labelled as Te$_2$, is set at the origin. The two Bi atoms are equivalent and located 
at $\pm \mu {\bf{c}}$. Here $\nu$ and $\mu$
are dimensionless parameters
and ${\bf{c}}={\bf{a}}_1+{\bf{a}}_2+{\bf{a}}_3$ 
is parallel to the trigonal axis (C$_3$ axis). The rhombohedral structure is depicted
in Fig. \ref{structure_fig}(a).
Alternatively,
the structure can be viewed as an hexagonal structure depicted in Fig. \ref{structure_fig}(b) and 
spanned by the following three lattice vectors
\begin{equation}\label{hex_cell}
{\bf{a}}_{\perp, 1}= {\bf{a}}_1 - {\bf{a}}_2 ;
~{\bf{a}}_{\perp, 2}= {\bf{a}}_2 - {\bf{a}}_3;
~{\bf{a}}_{\parallel}={\bf{c}},
\end{equation} 
where $a_{\perp, 1}=a_{\perp, 2} \equiv a_\perp$. The lattice
cell parameters of the two structures are related to each other by the following relations
\begin{equation}
\begin{array}{lcl}
a_\perp=2 a \sin\left(\frac{\alpha}{2}\right) & & a=\frac{1}{3} \sqrt{3 a_\perp^2+ a_\parallel^2}              \\
a_\parallel=a\sqrt{3+6\cos(\alpha)} &  &  \sin\left[\frac{\alpha}{2}\right]=\frac{3}{2} a_\perp/\sqrt{3 a_\perp^2+ a_\parallel^2.}
\end{array}
\end{equation}
The hexagonal structure is easier to visualize as compared to the rhombohedral
structure and can be viewed as made of three quintuple layers Te$_1$-Bi-Te$_2$-Bi-Te$_1$.
However, it's worth noting that all the calculations have been performed using 
the rhombohedral structure because it contains three times atoms less than the hexagonal structure.

The calculated LDA lattice parameters including spin-orbit coupling (SOC) are displayed 
in table \ref{lattice_tab} and compared to the lattice parameters obtained 
from X-ray diffraction experiments performed at 300 K\cite{francombe_1958} as well as to other
existing theoretical results\cite{luo_2012}. Our calculated lattice parameters 
$a_{\parallel,0}$ and $a_{\perp,0}$ are slightly larger than those obtained 
by Luo {\it{et al}}\cite{luo_2012}. Such a difference might be due to a different choice of the
pseudo-potentials and/or to a different implementation of the electronic structure calculation.
We also observe that our calculated $a_{\parallel,0}$ and $a_{\perp,0}$ are respectively 
underestimated from  2.6 \% and 0.3 \% with respect to the experimental values\cite{francombe_1958}.
Such an underestimation for $a_{\parallel,0}$ might be ascribed to thermal expansion effects ignored
in our calculations and to long range effects such 
as Van der Waals interactions not captured by the LDA exchange-correlation functional. 
It's worth considering the three relevant distances between successive planes perpendicular to
 the stacking axis (trigonal axis).
When considering
the fully relaxed structure, the two shortest distances $d_{Bi-Te_1}= 1.75$ \AA~ 
and $d_{Bi-Te_2}= 2.0$ \AA~ 
are in fairly good agreement with $d_{Bi-Te_1}^{~expt}= 1.74$ \AA~  and
$d_{Bi-Te_2}^{~expt}= 2.03$ \AA~
whereas the largest distance $d_{Te_1-Te_1}= 2.40$ \AA~
is strongly underestimated with respect to $d_{Te_1-Te_1}^{~expt}= 2.61$ \AA~ 
in accordance with the fact that
Van der Waals interactions between Te$_1$ atoms are expected to be significant. In 
contrast, the interplane distances $d_{Bi-Te_1}= 1.75$ \AA~, $d_{Bi-Te_2}= 2.04$ \AA~ and 
$d_{Te_1-Te_1}= 2.59$ \AA~ are almost correctly predicted when imposing the experimental
lattice parameters and relaxing the internal coordinates $\mu$ and $\nu$.
All the forthcoming calculations have been performed for the 
experimental lattice parameters\cite{francombe_1958}
and for the relaxed internal coordinates shown in table \ref{lattice_tab}. 

\section{ELECTRONIC STRUCTURE AT THE LDA LEVEL}\label{electronic_structure}
\begin{figure}[!htbp]
\begin{center}
\vskip1.2truecm
\includegraphics[angle=0, scale=0.5]
{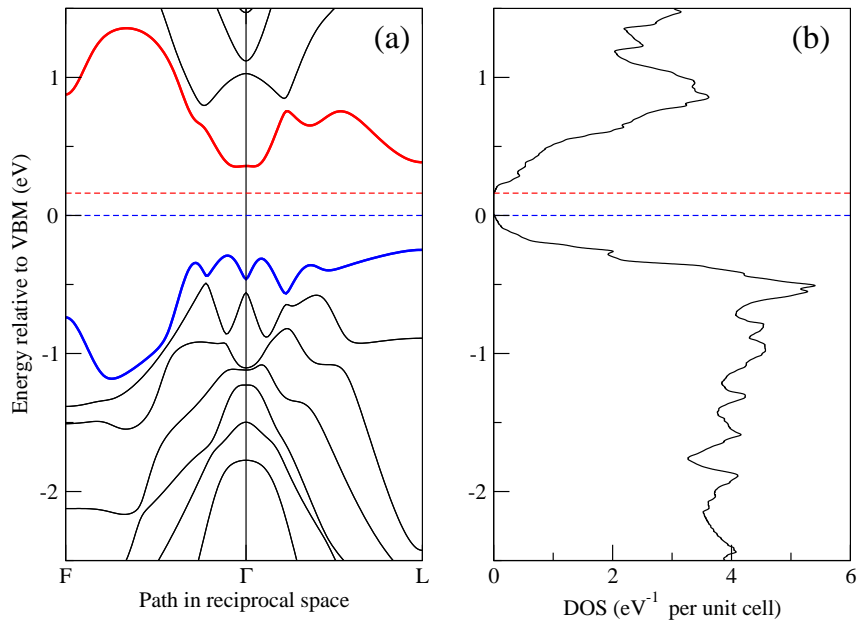}
\end{center}
\caption{(a) LDA bandstructure of Bi$_2$Te$_3$ 
including spin-orbit coupling (SOC) calculated along 
F-$\Gamma$-L where F$=(0, 0.5, 0.5)$ and L$=(0, 0.5, 0)$ in units of the reciprocal lattice
vectors. The last valence bands and first conduction bands 
are respectively colored in blue and red. 
(b) Density of states (DOS) per unit cell computed for 
a $256\times 256 \times 256$ grid of kpoints with a gaussian smearing of 10 meV. 
In both pannels, the valence band maximum is set at zero and displayed as a blue
dashed line while the conduction band minimum is displayed as a red dashed line. The
conduction bands are shifted towards higher energies from 120 meV to mimic the self-energy
corrections.
}
\label{BS_fig}
\end{figure}

\begin{figure}[!htbp]
\begin{center}
\includegraphics[angle=0, scale=0.4]
{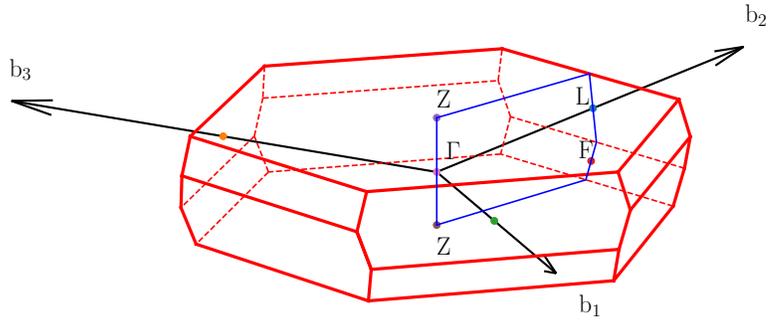}
\end{center}
\caption{
First Brillouin zone where the $x=0$ mirror plane (in blue) has been highlighted and 
where four high symetry ${\bf{k}}$-points, labelled $\Gamma$, $Z$, $L$ and $F$ are shown.
The $x=0$ mirror plane is spanned by the reciprocal lattice vectors ${\bf{b}}_2$ 
($\Gamma-L$ direction) and ${\bf{b}}_1+{\bf{b}}_2+{\bf{b}}_3$ ($\Gamma-Z$ direction)
and the two other mirror planes (not shown) are obtained by applying $C_3$ rotations 
around the trigonal axis to the $x=0$ mirror plane.
}
\label{bz_fig}
\end{figure}

The electronic structure of Bi$_2$Te$_3$ has received a lot of attention and has been
studied either at the DFT level\cite{luo_2012, wang_2007} 
or at the GW level\cite{kioupakis_2010, yazyev_2012, aguilera_2013, nechaev_2013}.
As shown in Fig. \ref{BS_fig}, the band structure, even at the LDA level, is rather complex. 
Indeed, the spin-orbit 
coupling (SOC) produces a band inversion in the vicinity of the $\Gamma$ point 
and shifts the band extremas from the high symmetry lines. Our calculated band extremas
are found in the three mirror planes in agreement with other 
theoretical studies\cite{luo_2012, wang_2007, kioupakis_2010, nechaev_2013}.
As two extremas are found in each mirror plane, the multiplicity of our calculated
extremas is 6 as suggested by Shubnikov-de Haas 
experiments\cite{kohler1_1976, kohler2_1976} and Landau level 
spectroscopy\cite{mohelsky_2020}. The extremas of the valence band and conduction band 
in the $x=0$ mirror plane depicted in Fig. \ref{bz_fig} 
are respectively found at $\pm (0.349, 0.523, 0.349)$ 
and $\pm (0.542, 0.647, 0.542)$ in units of the reciprocal lattice vectors while
these extremas are respectively located at $\pm (0.37, 0.54, 0.37)$ 
and $\pm (0.58, 0.68, 0.58)$ in Ref. \cite{kioupakis_2010}. 
Furthermore, the
calculated LDA indirect gap is $\sim 42$ meV while the minimum direct gap
is $\sim 91$ meV. The indirect gap differs from the LDA gap $\sim 87$ meV reported
in Ref. \cite{kioupakis_2010}. Such a difference might be ascribed to the lattice cell
parameters and/or internal coordinates that are expected to play a significant role
on both the positions of the extremas and the values of the gaps\cite{nechaev_2013}.
Optical measurements performed at 10 K led to a probably indirect band gap of
$150\pm 20$ meV\cite{thomas_1992} and a direct gap 
of $220\pm 20$ meV\cite{thomas_1992} confirmed by more 
recent ellipsometry measurements\cite{dubroka_2017}. The discrepancy between the theoretical
and experimental results is noticeable and reflects the fact that the LDA eigenvalues 
can not be interpreted as quasiparticle energies. More sophisticated approaches based on the
evaluation of the self-energy within 
the GW approximation\cite{hybertsen_1986, onida_2002, arnaud_2000, lebegue_2003}, 
are needed to provide a better description of quasiparticle 
properties. All the GW calculations\cite{kioupakis_2010, yazyev_2012, aguilera_2013, nechaev_2013}
produce band gaps in better agreement with experiments. It's worth highlighting that
an accurate description of the quasiparticle bandstructure near the $\Gamma$ point 
necessitates to take into account the non diagonal elements 
of the self-energy operator\cite{aguilera_2013}.
The self-energy corrections, while less than 200 meV for one band and one kpoint\cite{yazyev_2012}, 
are highly non trivial. Indeed, the rigid shift scissor approximation is not valid
for the highest valence band and lowest conduction band\cite{yazyev_2012, aguilera_2013},
especially in the vicinity of the $\Gamma$ point where the band inversion occurs and 
where the direct LDA band gaps are reduced\cite{aguilera_2013}. However, as this approximation
seems to be well founded for other bands, we shift all conduction bands from 120 meV with respect
to the valence bands to reproduce both the experimental indirect and direct gaps and keep
in mind that we only achieve a poor description of the self-energy corrections needed to
compute the optical properties beyond the standard random-phase approximation (RPA).

\section{OPTICAL PROPERTIES AT THE RPA LEVEL}\label{optical_properties}
Within the RPA\cite{adler_1962, wiser_1963}, 
the dielectric matrix  for a given ${\bf q}$ wavevector reads:
\begin{eqnarray}\label{adler_wiser}
\epsilon_{{\bf G},{\bf G}^{\prime}}({\bf q},\omega) &
= &\delta_{{\bf G},{\bf G}^{\prime}}
        - \frac{e^2}{4\pi\epsilon_0}\frac{4 \pi}{v |{\bf q}+{\bf G}| |{\bf q}+{\bf G}^{\prime}|}
        \frac{1}{N}\sum_{{\bf k},n,m}\frac
        {f_{n, {\bf k}-{\bf q}} - f_{m,{\bf k}}}
        {\epsilon_{n{\bf k}-{\bf q}}-\epsilon_{m{\bf k}} +\hbar\omega + i\delta} \nonumber \\
 & &   \times \langle n{\bf k}-{\bf q}|e^{-i({\bf q}+{\bf G}).{\bf r}}|m{\bf k}\rangle
      \langle m{\bf k}|e^{i({\bf q}+{\bf G}^{\prime}).{\bf r}}|n{\bf k}-{\bf q}\rangle
\end{eqnarray}
where $({\bf G},{\bf G}^{\prime})$ is a couple of reciprocal lattice vectors, 
$\epsilon_{m{\bf k}}$ are the Kohn-Sham energies for band $m$ and wave vector ${\bf k}$,
$f_{m,{\bf k}}\equiv f(\epsilon_{m{\bf k}})$ are
the occupation numbers within the Fermi-Dirac distribution $f$ at temperature $T$, 
$v$ is the unit cell volume and $N$ is the number of kpoints included in the summation.  
In this expression, the time dependence of the field 
was assumed to be $e^{-i\omega t}$ and the small
positively defined constant $\delta$ guarantees that the matrix elements of 
$\epsilon(\omega)$ are analytic functions in the upper half plane. The macroscopic dielectric tensor
$\overline{\overline{\epsilon}}(\omega)$ is a $3\times3$ matrix given by:
\begin{equation}\label{macroscopic_tensor}
\hat{{\bf q}}^\top ~\overline{\overline{\epsilon}}(\omega) ~\hat{{\bf q}} = 
1/\lim_{{\bf q} \to 0} \epsilon^{-1}_{{\bf{0}}, {\bf{0}}} =
\lim_{{\bf q} \to 0}\epsilon_{0,0}({\bf q},\omega)
                -\lim_{{\bf q} \to 0} \sum_{{\bf G},{\bf G}^{\prime}\neq {\bf{0}}}
                 \epsilon_{0,{\bf G}}({\bf q},\omega)
                 \epsilon^{-1}_{{\bf G},{\bf G}^{\prime}}({\bf q},\omega)
                 \epsilon_{{\bf G}^{\prime},0}({\bf q},\omega),
\end{equation}
where $\hat{{\bf q}}$ is a unitary vector along the wavevector ${\bf q}$. 
Neglecting the local field (LF)  effects amounts to assuming 
that the dielectric matrix is diagonal, i.e.
to neglecting the second term in Eq. \ref{macroscopic_tensor}.
Given the hexagonal symmetry of Bi$_2$Te$_3$, $\overline{\overline{\epsilon}}(\omega)$
is diagonal with only two independent elements, namely $\epsilon_\perp(\omega)$ 
and $\epsilon_\parallel(\omega)$ that are respectively obtained for $\hat{{\bf q}}$
perpendicular and $\hat{{\bf q}}$ parallel to the trigonal axis. The dielectric matrix
defined in Eq. \ref{adler_wiser} is computed with the YAMBO code\cite{yambo_2009, yambo_2019}
which requires the ground state electronic structure computed with 
the ABINIT code\cite{gonze_2009, gonze_2016}.
Indeed, the Kohn-Sham eigenvalues 
$\epsilon_{m{\bf k}}$ and eigenvectors $|m {\bf{k}}\rangle$ are key ingredients to 
compute the dielectric function at the RPA level. From a practical point of view,
we used a $64\times 64 \times 64$ kpoints grid (22913 irreducible kpoints) and
included 28 valence bands and 34 conduction bands in our calculations to converge the
dielectric function without LF effects. The LF effects are included by taking into account
the second term in Eq. \ref{macroscopic_tensor}, where a $302 \times 302$ body matrix
$\epsilon_{{\bf G},{\bf G}^{\prime}}({\bf q}\to {\bf{0}},\omega)$ with 
${\bf G}, {\bf G}^{\prime} \neq {\bf{0}}$, ensures the convergence of the macroscopic 
dielectric function.
\begin{figure}[!htbp]
\begin{center}
\vskip1.2truecm
\includegraphics[angle=0, scale=0.5]
{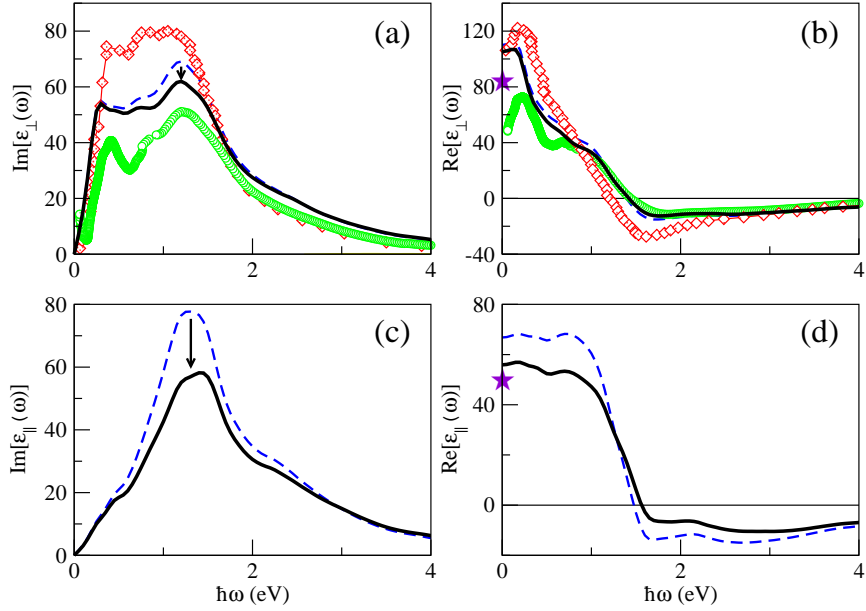}
\end{center}
\caption{Imaginary and real part of the  dielectric
function of Bi$_2$Te$_3$ computed at the RPA level with LF effects (thick black lines) and 
without LF effects (blue dashed lines) for an electric field perpendicular 
to the trigonal axis (upper pannels) and parallel to the trigonal axis
(lower pannels) as a function of the photon energy $\hbar \omega$ (in eV). All the conduction
bands have been shifted from 120 meV with respect to the valence bands in order to roughly 
reproduce the self-energy corrections. The imaginary and real part of the dielectric function
obtained from ellipsometry measurements\cite{dubroka_2017} and from a Kramers Kronig analysis
of reflectivity measurements\cite{Greenaway_1965} are respectively displayed as green circles
and red diamonds in the upper pannels. The two vertical arrows in pannels (a) and (c) show
how the imaginary part of the dielectric function evolves when LF effects are included.
The stars in pannel (b) and (d) denote the experimental clamped-nuclei 
static dielectric constants extracted from the IR reflectance 
spectras measured at 300 K\cite{richter_1977}.
}
\label{Im_Re_epsilon_fig}
\end{figure}

Both the imaginary part and the
real part of the dielectric function $\epsilon_{\perp, \parallel}(\omega)$ 
computed with LF effects (thick black lines) and without LF effects (blue dashed lines) 
at $T=300$ K are displayed in Fig. \ref{Im_Re_epsilon_fig} and compared 
to the experimental results\cite{dubroka_2017, Greenaway_1965} 
shown as circles and diamonds in the upper pannels of Fig. \ref{Im_Re_epsilon_fig}. 
The imaginary part of the dielectric function computed without LF effects 
for an electric field ${\bf{E}}\perp {\bf{c}}$ (see Fig. \ref{Im_Re_epsilon_fig}(a))
displays two peaks respectively located at 0.32 eV and 1.21 eV while 
the experimental peaks\cite{dubroka_2017} are respectively located at 0.42 eV and 1.22 eV.
Such a difference might be related to our crude implementation of the self-energy corrections
(the conduction bands have been shifted from 120 meV with respect to the valence bands) or
to excitonic effects neglected at our level of approximation\cite{onida_2002, arnaud_2001, arnaud_2006}. It's important to point out
that the peak located at $0.32$ eV is sensitive to the temperature $T$ and disappears 
at $T=300$ K when the conduction bands are not shifted, which can be explained by the 
unphysical thermal excitation of electrons from the highest valence band 
to the lowest conduction band arising
from the tiny indirect band gap $\sim 42$ meV at the LDA level. 
We also observe that the LF effects 
do not affect the positions of the peaks but slightly reduce the intensity of the peak 
located at 1.21 eV. The situation is clearly different for ${\bf{E}}\parallel {\bf{c}}$
(see Fig. \ref{Im_Re_epsilon_fig}(c)). Indeed, the imaginary part of the dielectric function
computed without LF effects displays one peak located at 1.31 eV which is shifted to 1.43 eV
when LF effects are included. Furthermore, the intensity of this peak is decreased by
$\sim 25$ \% when LF effects are included, suggesting that LF effects are crucial to
predict the optical properties for ${\bf{E}}\parallel {\bf{c}}$. This observation is somehow 
expected as the LF effects, vanishing for an homogeneous electron gas, 
are as large as the degree of inhomogeneity in the charge density
increases\cite{arnaud_2001}. Indeed, the inhomogeneity in the charge density is undeniably 
strong along the trigonal axis, namely the stacking axis of the quintuple layers. 

We observe a discrepancy between theory and experiment
as our calculated imaginary part of the dielectric function for ${\bf{E}}\perp {\bf{c}}$ 
 (see Fig. \ref{Im_Re_epsilon_fig}(a)) falls in between the two experimental 
results\cite{dubroka_2017, Greenaway_1965}. Importantly,
the two experimental results only coincide above 3 eV, where the anisotropy in the optical
response starts to be negligible as shown in our calculations with or without LF effects.
The experimental results of Ref.\cite{Greenaway_1965} are deduced from a Kramers-Kronig (KK)
transformation of the near-normal incidence reflectivity spectra for ${\bf{E}}\perp {\bf{c}}$.
Thus, both the real and imaginary part of the dielectric function might be erroneous because
they are very sensitive to small errors in the reflectivity data and in the high energy
extrapolation. The experimental results obtained 
from ellipsometry measurements\cite{dubroka_2017} might also be in error as the authors
assumed that they only probe the in-plane component
of the dielectric function ($\epsilon_\perp$) in the oblique
geometry of their measurements while they also probed the out of plane component of the
dielectric function ($\epsilon_\parallel$), introducing some errors in the determination 
of $\epsilon_\perp$ as our calculations show that the anisotropy in the optical response
up to 3 eV is far from being negligible. Therefore, it's difficult to draw a definitive
conclusion as two measurements do not lead to the same results, highlighting the
need for new experimental studies. It's also important to point out that excitonic 
effects, neglected at the RPA level, might change the calculated dielectric function.
However, excitonic effects are expected to be small because of the large dielectric screening.
Indeed, our clamped-nuclei static dielectric constants $\epsilon_{\infty, \perp}$ 
and $\epsilon_{\infty, \parallel}$ with (without) LF effects are respectively equal to
105 (110) and 56 (67), while the experimental static dielectric constants 
infered from infrared spectroscopy measurements\cite{richter_1977} and shown as stars in
Fig. \ref{Im_Re_epsilon_fig}(b) and \ref{Im_Re_epsilon_fig}(d), are respectively equal to 85
and 50. 
Such a large contribution of the electrons to the static dielectric constants
can be understood from the Kramers Kronig
transformation that relates the real part of the dielectric function to the imaginary 
part of the dielectric function. Indeed, we have
\begin{equation}\label{kramers-kronig}
\epsilon_{\infty, \perp, \parallel}= \frac{2}{\pi} 
\int_0^\infty d\omega ~\frac{\textrm{Im}\left[\epsilon_{\perp, \parallel}(\omega)\right]   }{\omega}.
\end{equation}
As $\textrm{Im}\left[\epsilon_{\perp}(\omega) \right] > 
\textrm{Im}\left[\epsilon_{\parallel}(\omega) \right]$ for $\hbar \omega$ below 1.4 eV
, we can conclude that $\epsilon_{\infty, \perp} > \epsilon_{\infty, \parallel}$ 
since the anisotropy in the optical response is weak above 1.4 eV when LF effects are included.
Furthermore, $\epsilon_{\infty, \perp}$ and $\epsilon_{\infty, \parallel}$ are very large
because of strong interband transitions starting from 210 meV (direct LDA band gap increased
from 120 meV) that strongly contribute to the static dielectric constant because of the $1/\omega$
weighting  factor in the integrand of Eq. \ref{kramers-kronig}.

Interestingly, as shown in Fig. \ref{Im_Re_epsilon_fig}(b) and Fig. \ref{Im_Re_epsilon_fig}(d), 
the real part of $\epsilon_{\perp}$ ($\epsilon_{\parallel}$) including
LF effects crosses the zero axis around 1.42 eV (1.56 eV) and becomes negative up 
to 6 eV. Thus, plasmonic applications, as nicely demonstrated in Ref.\cite{zhao_2015}, 
can be envisionned
but with the inconvenient that the imaginary part of the dielectric function is far
from being negligible in this spectral range, especially in the visible range. 
Our calculations also show that Bi$_2$Te$_3$ behaves as an hyperbolic material in 
a very narrow range of energy between 1.42 and 1.56 eV where the permittivity components
in different directions have opposite sign ($\epsilon_{\perp} \cdot \epsilon_{\parallel} <0$).
Therefore, our theoretical results contradict the ellipsometry measurements 
of Ref.\cite{esslinger_2014},
where the authors claimed that Bi$_2$Te$_3$ is a natural hyperbolic material in the visible
range. Such a discrepancy might be ascribed to the fact that the real part of the dielectric
function for ${\bf{E}} \parallel {\bf{c}}$ is incorrectly measured.

We also computed the normal incidence reflectivity according to
\begin{equation}\label{def_R_calc}
R_{\perp, \parallel}(\omega)=\left|\frac{n_{\perp, \parallel}(\omega)-1}
{n_{\perp, \parallel}(\omega)+1}   \right|^2,
\end{equation}
where $n_{\perp, \parallel}(\omega)=\sqrt{\epsilon_{\perp, \parallel}(\omega)}$ is 
the optical index calculated with or without LF effects 
for $\mathbf{E} \perp \mathbf{c}$ or $\mathbf{E} \parallel \mathbf{c}$. 
\begin{figure}[!htbp]
\begin{center}
\vskip1.2truecm
\includegraphics[angle=0, scale=0.5]
{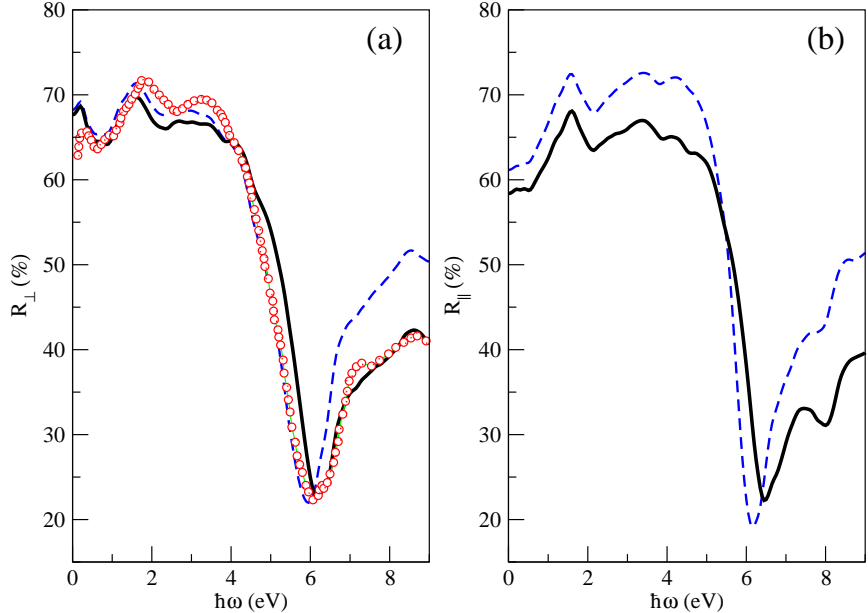}
\end{center}
\caption{
Computed optical reflectivity with (thick black lines) and without 
(thin blue dashed lines) LF effects for $\mathbf{E} \perp \mathbf{c}$ (left pannel) 
and $\mathbf{E} \parallel \mathbf{c}$ (right pannel) compared to the near-normal incidence reflectance
spectra (open circles) measured at 300 K for $\mathbf{E} \perp \mathbf{c}$\cite{Greenaway_1965}.  }
\label{Optical_Reflectivity_fig}
\end{figure}
As shown in Fig. \ref{Optical_Reflectivity_fig}(a),
the overall agreement between theory and experiment\cite{Greenaway_1965} 
for $\mathbf{E} \perp \mathbf{c}$ is fairly good when LF effects are included. 
It's worth remarking that LF effects
do not change the reflectivity spectra very much between 0 eV and 4 eV as can be inferred
from Fig. \ref{Im_Re_epsilon_fig}(a) and Fig. \ref{Im_Re_epsilon_fig}(b) 
but lead to a strong decrease of the calculated
reflectivity from 6 to 9 eV bringing the calculated spectra in closer agreement with
the experimental spectra\cite{Greenaway_1965}.
The situation is more contrasted for $\mathbf{E} \parallel \mathbf{c}$ as LF effects
are noticeable everywhere. As shown 
in Fig. \ref{Optical_Reflectivity_fig}(b), the calculated reflectivity with LF effects
is much smaller than the calculated reflectivity without LF effects almost everywhere
up to 9 eV, except for a narrow range of energy between 5.5 eV and 6.4 eV, where the
reflectivity displays a dip.

\section{FREQUENCIES AT THE ZONE CENTER}\label{frequencies} 
As the primitive cell contains five atoms, there are 15 lattice dynamical modes at
the Brillouin zone center, three of which are acoustic modes. Group theory classifies the
remaining 12 optical modes into 2 A$_{1g}$ (R), 2 E$_g$ (R), 2 A$_{2u}$ (IR)
and 2 E$_{u}$ (IR) modes, where R and IR refer to Raman and infrared
active modes respectively.
We computed the zone center frequencies within a DFPT approach\cite{gonze_1997} 
for the experimental structure at 300 K\cite{francombe_1958}, where only the internal 
coordinates have been relaxed (see table \ref{lattice_tab}).
\begin{table}[!hbt]
\caption{Experimental and theoretical values of Raman and infrared
phonon frequencies for Bi$_2$Te$_3$ in units of cm$^{-1}$. 
Experimental Raman\cite{richter_1977, wang_2013}
and infrared values\cite{richter_1977} were measured at 300 K. 
The values within parentheses correspond to the frequencies in THz and the most recent experimental
results\cite{wang_2013} are underlined.}
\label{frequency_tab}
\begin{tabular}{lccc}
\hline
Symmetry & \multicolumn{2}{c}{Experiment}  & Theory  \\
\hline
E$_g^1$  & -         &   \underline{37} (\underline{1.11})  &  36.81 (1.10)   \\
A$_{1g}^1$  & 62.5 (1.87)  & \underline{62}  (\underline{1.86})  & 57.25  (1.71) \\
E$_g^2$  & 103    (3.09)   & \underline{102}  (\underline{3.06})  &  97.96 (2.93) \\
A$_{1g}^2$  & 134 (4.02)   & \underline{135}   (\underline{4.05}) & 129.25 (3.87) \\
E$_u^1$   & $50\pm2$ (1.50 $\pm 0.06$) &        & 54.25  (1.62) \\
A$_{2u}^1$  & $94\pm4$ (2.82 $\pm 0.12$) &      & 93.23  (2.79)  \\
E$_u^2$   & $95\pm5$    (2.85 $\pm 0.15$) &     & 91.75 (2.75) \\
A$_{2u}^2$   & $120\pm5$  (3.60 $\pm 0.15$) &   &  120.25 (3.60) \\
\hline
\end{tabular}
\end{table}
As shown in Table \ref{frequency_tab},
the overall agreement between theory and experiment\cite{richter_1977, wang_2013}
is satisfactorily 
and the largest discrepancy occurs for the A$_{1g}^1$ and
E$_u^1$ frequencies that are respectively underestimated and overestimated from 8\% with
respect to the experimental frequencies. Such a difference might be ascribed to 
anharmonic effects not captured in our calculations\cite{yang_2020}. 

\section{BORN EFFECTIVE CHARGES}\label{Born_Charges}
We used a finite electric field approach\cite{souza_2002} in order to compute the
Born effective charge tensors. The force along $\beta$ acting on atom $p$ reads
\begin{equation}\label{def_charge}
F_p^\beta = e \sum_\alpha Z^*_{p, \alpha, \beta} E^\alpha,
\end{equation}
where $E^\alpha$ is the component along $\alpha$ of the macroscopic electric field and
$Z^*_{p, \alpha, \beta}$ is the Born effective charge tensor defined for each atom $p$.
Given the hexagonal symmetry of Bi$_2$Te$_3$, the tensor is diagonal and reduces to 
two values $Z^*_{p, \perp}=Z^*_{p, 1, 1}=Z^*_{p, 2, 2}$ and $Z^*_{p, \parallel}=Z^*_{p, 3, 3}$ 
for each atom. Using Eq. \ref{def_charge}, we conclude that
\begin{equation}
\left\{
\begin{array}{ll}
F_p^\beta = Z^*_{p, \perp} E^\beta & ~\textrm{when}~ \beta=1, 2  \\
F_p^\beta = Z^*_{p, \parallel} E^\beta & ~\textrm{when}~ \beta=3  \\
\end{array}\right.
\end{equation}
Thus, we computed the forces acting on each atom for an electric field perpendicular
and parallel to the trigonal axis whose amplitude is varied from $-10^{-6}$ a.u
to $10^{-6}$ with a step of $0.5\times 10^{-6}$ a.u (2.57 kV.cm$^{-1}$).
\begin{figure}[!hbt]
\begin{center}
\vskip0.5truecm
\includegraphics[angle=0, scale=0.5]
{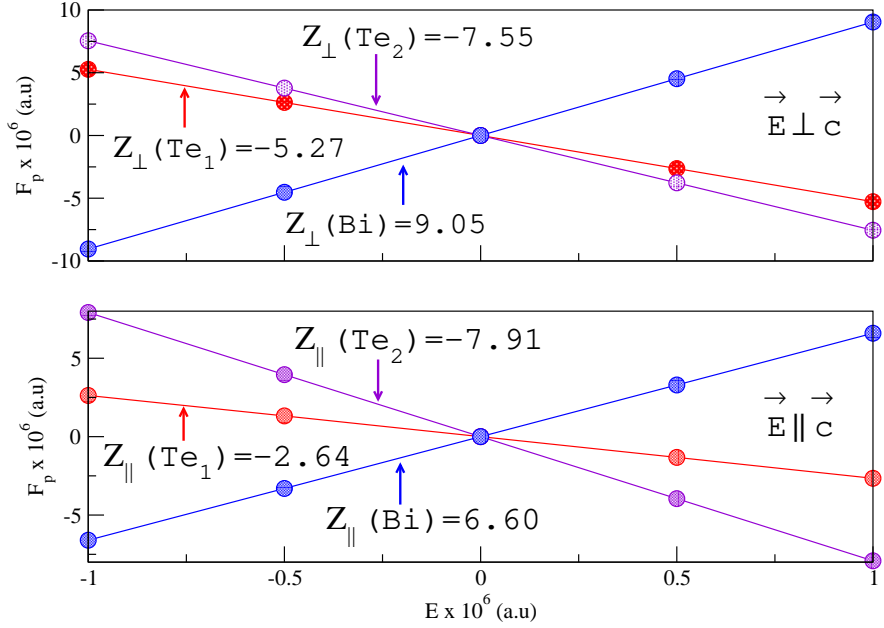}
\end{center}
\caption{
\label{Born-effective-charges-fig}
Forces $F_p$ (in a.u) as a function of the macroscopic electric field $E$ (in a.u)
applied perpendicular (upper pannel) and parallel (lower pannel) to the trigonal axis
for the experimental structure at 300 K\cite{francombe_1958}.
}
\end{figure}
As shown in Fig. \ref{Born-effective-charges-fig}, the forces $F_p$ acting on each atom $p$
vary linearly with the amplitude of the macroscopic electric field $E$. A linear fitting
provides the Born effective charges that are gathered in table \ref{effective_charge_tab}. 
\begin{table}
\caption{Calculated Born effective charges 
for the two equivalent Bi atoms, the two equivalent Te$_1$ atoms and the
Te$_2$ atom underlying the five-point basis. The degree of anisotropy in the
Born effective charges is quantified by Z$_\parallel^*/$Z$_\perp^*$.
} 
\label{effective_charge_tab}
\begin{tabular}{lccc}
\hline
   & Bi & Te$_1$ & Te$_2$ \\
\hline
Z$_\parallel^*$  & 6.60 & -2.64 &  -7.91 \\     
Z$_\perp^*$      & 9.05 & -5.27  &  -7.55   \\ 
Z$_\parallel^*/$Z$_\perp^*$   & 0.729 & 0.500 & 1.047    \\ 
\hline
\end{tabular}
\end{table}
It's worth pointing out that the acoustic sum rule 
$\sum_p Z^*_{p, \perp~\textrm{or}~\parallel}$ is fulfilled to within $10^{-5}$ suggesting that
our calculations are well converged. 
The effective charges are large 
and differ strongly from what can be expected for fully ionic configurations corresponding
to closed shell ions ($+3$ for Bi and $-2$ for Te). However, it's worth reminding that
effective charges are dynamic only and reflect the effect of covalency\cite{lee_2003}.
Interestingly, the Born effective charges have
surprisingly different values for the Te$_1$ and Te$_2$ atoms and show large anisotropy
for the Te$_1$ atoms that form strongly covalent bonds with the Bi atoms as the distance
between the Te$_1$ and Bi atoms is the shortest one. The Born effective charges of Bi 
atoms also display some degree of anisotropy. 
From a physical point of view, the two
Bi atoms and the two Te$_1$ atoms behave respectively as anisotropic cations and anions,
while the Te$_2$ atom behaves as an isotropic anion. 

\section{INFRARED DIELECTRIC TENSOR}\label{dielectric_tensor} 
We consider a unit cell of volume $v$ containing $n$ atoms. The total energy
of the unit cell $E_{tot}$, when a macroscopic electric field $\bf{E}$ is applied,
can be taylor expanded as:
\begin{equation}
E_{tot}=
E_0 + \frac{1}{2} \sum_{p, p^\prime, \alpha, \alpha^\prime} 
u_p^\alpha \frac{\partial^2 E_{tot}}{\partial u_p^\alpha \partial u_{p^\prime}^{\alpha^\prime}} 
u_{p^\prime}^{\alpha^\prime}
+\sum_{p, \alpha, \beta} 
\frac{\partial^2 E_{tot}}{\partial u_p^\alpha \partial E^\beta} u_p^\alpha E^\beta
+\frac{1}{2} \sum_{\alpha, \beta} 
\frac{\partial^2 E_{tot}}{\partial E^\alpha \partial E^\beta} E^\alpha E^\beta,
\end{equation}
where $u_p^\alpha$ is 
the displacement of atom $p$ ($p \in \{1, \cdots, n\}$) along the direction $\alpha$
and $E^\beta$ the component of the macroscopic field along $\beta$ with
$\alpha, \beta \in \{1,2,3\}$. We can introduce the elastic constants 
$C_{p, p^\prime}^{\alpha, \alpha^\prime}$, the dimensionless  
Born effective charge tensor $Z^*_{p, \beta, \alpha}$ and the dimensionless
electronic tensor susceptibility $\chi_{\alpha, \beta}$, respectively defined by
\begin{equation}
C_{p, p^\prime}^{\alpha, \alpha^\prime} = 
\frac{\partial^2 E_{tot}}{\partial u_p^\alpha \partial u_{p^\prime}^{\alpha^\prime}},~
Z^*_{p, \beta, \alpha}=-\frac{1}{e} \frac{\partial^2 E_{tot}}{\partial u_p^\alpha \partial E^\beta},~
\textrm{and}~
\chi_{\alpha, \beta} = -\frac{1}{v \epsilon_0} 
\frac{\partial^2 E_{tot}}{\partial E^\alpha \partial E^\beta}.
\end{equation}
Thus, the total energy reads
\begin{equation}\label{total_energy}
E_{tot}=
E_0 + \frac{1}{2} \sum_{p, p^\prime, \alpha, \alpha^\prime} 
u_p^\alpha C_{p, p^\prime}^{\alpha, \alpha^\prime} u_{p^\prime}^{\alpha^\prime}
-e \sum_{p, \alpha, \beta} Z^*_{p, \beta, \alpha} u_p^\alpha E^\beta
-\frac{1}{2} v \epsilon_0 \sum_{\alpha, \beta} \chi_{\alpha, \beta} 
E^\alpha E^\beta,
\end{equation}
leading to a force (See Eq. \ref{def_charge}) exerted on atom $p$ by the macroscopic elecric 
field when the atoms are held at their equilibrium position.
From Eq. \ref{total_energy}, we get the component along $\beta$ of the polarization
\begin{equation}
P^\beta=-\frac{1}{v} \frac{\partial E_{tot} }{\partial E^\beta }= P^{\beta, ion} + P^{\beta, el},
\end{equation}
where the ionic contribution to the polarization reads
\begin{equation}\label{pol_ionic}
P^{\beta, ion} = \frac{e}{v} \sum_{\alpha, p} Z^*_{p, \beta, \alpha} u_p^\alpha,
\end{equation}
and the electronic contribution reads
\begin{equation}\label{pol_el}
P^{\beta, el} = \epsilon_0 \sum_{\alpha} \chi_{\beta, \alpha} E^\alpha.
\end{equation}

Using Eq. \ref{total_energy}, the equation of motion for atom $p$ of mass $M_p$ reads
\begin{equation}\label{eq_motion}
M_p \ddot{u}_p^\alpha = - \frac{\partial E_{tot}   }{\partial u_p^\alpha  } =
- \sum_{p^\prime, \alpha^\prime} 
C_{p, p^\prime}^{\alpha, \alpha^\prime} u_{p^\prime}^{\alpha^\prime}
+ e \sum_\beta Z^*_{p, \beta, \alpha} E^\beta.
\end{equation}
When $\mathbf{E}=\mathbf{0}$, we can plug
\begin{equation}
u_p^\alpha = \sqrt{\frac{M}{M_p}} \epsilon_p^\alpha \exp[-i \omega t ],
\end{equation}
where $M=\sum_p M_p$,  in Eq. \ref{eq_motion} 
and solve the following eigenvalue problem
\begin{equation}\label{eigen_problem}
\sum_{p^\prime, \alpha^\prime} D_{p, p^\prime}^{\alpha, \alpha^\prime}
\epsilon_{p^\prime}^{\alpha^\prime}(\lambda) = \omega_\lambda^2 \epsilon_{p}^{\alpha}(\lambda),
\end{equation}
where $D_{p, p^\prime}^{\alpha, \alpha^\prime}=C_{p, p^\prime}^{\alpha, \alpha^\prime}/\sqrt{M_p M_{p^\prime}}$ is the dynamical matrix at the zone center. Here $\omega_\lambda$ 
and $\epsilon_{p}^{\alpha}(\lambda)$ are respectively the frequency and the displacement of
atom $p$ along $\alpha$ for the mode $\lambda$, 
where $\lambda \in \{1, \cdots, 3n\}$.
The eigenvectors of Eq. \ref{eigen_problem} satisfy the orthogonality relations
\begin{equation}\label{orthogonality}
\sum_{p, \alpha} \epsilon_{p}^{\alpha}(\lambda) \epsilon_{p}^{\alpha}(\lambda^\prime) =
\delta_{\lambda, \lambda^\prime} \Longleftrightarrow 
\mathbf{\epsilon}(\lambda)^\top \mathbf{\epsilon}(\lambda^\prime) =
\delta_{\lambda, \lambda^\prime},
\end{equation}
since the zone center dynamical matrix is real and symmetric.

When $\mathbf{E}=\mathbf{E}_0 \exp(-i\omega t)$, where $\omega$ is the frequency 
of the electric field in the THz range, we can plug
\begin{equation}\label{u_def}
u_p^\alpha = \sqrt{\frac{M}{M_p}} z_p^\alpha \exp[-i \omega t ],
\end{equation}
into Eq. \ref{eq_motion}. Thereby, the following equation
\begin{equation}\label{eigen_problem2}
\omega^2 z_p^\alpha = \sum_{p^\prime, \alpha^\prime} D_{p, p^\prime}^{\alpha, \alpha^\prime}
z_{p^\prime}^{\alpha^\prime}
-\frac{e}{\sqrt{M M_p}} \sum_\beta Z^*_{p, \beta, \alpha} E_0^\beta  
\end{equation}
must be solved. As $\{ \mathbf{\epsilon}(\lambda), \lambda=1, \cdots, 3n \}$ forms
a complete basis set, we can seek a solution in the form
\begin{equation}\label{z_eq}
z_p^\alpha = \sum_\lambda A(\lambda) \epsilon_p^\alpha(\lambda),
\end{equation} 
where the coefficients $A(\lambda)$ have to be determined. Introducing Eq. \ref{z_eq}
in Eq. \ref{eigen_problem2} and using Eq. \ref{eigen_problem} leads to
\begin{equation}
\sum_{\lambda^\prime} A(\lambda^\prime) \left[ \omega^2 - \omega_{\lambda^\prime}^2 \right] \epsilon_p^\alpha(\lambda^\prime) =
-\frac{e}{\sqrt{M M_p}} \sum_\beta Z^*_{p, \beta, \alpha} E_0^\beta
\end{equation}
By multiplying this identity with $\epsilon_p^\alpha(\lambda)$, summing over $p$ and
$\alpha$ and using the orthogonality relation (Eq. \ref{orthogonality}), we obtain
\begin{equation}\label{def_A}
A(\lambda)=-\frac{e}{M} \frac{1}{\omega^2 - \omega_\lambda^2} \sum_{\beta^\prime} 
Z_{\beta^\prime}(\lambda) E_0^{\beta^\prime},
\end{equation} 
where the mode effective charge
\begin{equation}\label{def_X}
Z_{\beta} (\lambda) = \sum_{p, \alpha} \sqrt{\frac{M}{M_p}} \epsilon_p^\alpha(\lambda)
Z^*_{p, \beta, \alpha},
\end{equation} 
is a dimensionless quantity. Plugging Eq. \ref{def_A} in Eq. \ref{z_eq}, and using
the definition of ionic polarization (Eq. \ref{pol_ionic}) as well as Eq. \ref{u_def},
we obtain
\begin{equation}
P^{\beta, ion} (\omega) = \frac{e^2}{v} \frac{1}{M} \sum_{\beta^\prime, \lambda} 
\frac{ Z_\beta(\lambda) Z_{\beta^\prime}(\lambda)}{\omega_\lambda^2 - \omega^2 } E^{\beta^\prime}
\end{equation} 
By analogy with the definition of the electronic susceptibility (Eq. \ref{pol_el}),
we can define an ionic susceptibility 
\begin{equation}\label{Chi_def}
\chi^{ion}_{\beta, \beta^\prime  } (\omega)= \frac{e^2}{\epsilon_0 v} \frac{1}{M} \sum_{\lambda} 
\frac{ S_{\beta, \beta^\prime} (\lambda) }{\omega_\lambda^2- \omega^2 -2 i \omega \gamma_\lambda},
\end{equation} 
where $2 \gamma_\lambda$ is the full width at half maximum (FWHM) of the phonon peak and 
$S_{\beta, \beta^\prime} (\lambda)=Z_\beta(\lambda) Z_{\beta^\prime}(\lambda) $ 
is a dimensionless $3\times 3$ real symmetric matrix.
It's straightforward to show  using Eq. \ref{def_X} that
this matrix $S$ is zero for a gerade mode. Interestingly, $S$ might be non diagonal 
for degenerated modes like the $E_u^1$ mode at 1.62 THz or the $E_u^2$ mode at 2.75 THz that
can couple to an electric field perpendicular to the trigonal axis. However, the matrix
$S$ becomes diagonal when the polarization of the first $E_u$ mode is chosen along
one of the two fold axis (x-axis) while the polarization of the second $E_u$ mode lies
along one of the mirror plane (y-axis). Thus, we get:
\begin{equation}\label{chi_ionic_def2}
\chi^{ion}_{\perp, \parallel}(\nu) =
\frac{e^2}{4\pi^2 \epsilon_0 v} \frac{1}{M} \sum_{\lambda} 
\frac{ S_{\perp, \parallel} (\lambda) }{\nu_\lambda^2- \nu^2 - i \nu \gamma_\lambda/\pi},
\end{equation} 
where $S_\perp (\lambda)=\left[Z_1(\lambda)\right]^2~\textrm{or}~\left[Z_2(\lambda)\right]^2$,
$S_\parallel (\lambda)=\left[Z_3(\lambda)\right]^2$
(see Eq. \ref{def_X}) and where the summation can be restricted to the IR active modes. 
\begin{table}[!hbt]
\caption{Frequencies (in THz), oscillator strengths (dimensionless quantities)
and normalized atomic displacements (see Eq. \ref{orthogonality}) 
for the IR active modes. The calculations have been done for the experimental
lattice structure\cite{francombe_1958} using the values of the Born effective charges
reported in table \ref{effective_charge_tab}.
}
\label{osc_strength_tab}
\begin{tabular}{lccccccccc}
\hline
Symmetry & Frequency &\multicolumn{2}{c}{Oscillator strengths} &\multicolumn{5}{c}{Normalized atomic displacements} & Direction\\
         &           & $S_\perp$  & $S_\parallel$ &Te$_1$ & Bi & Te$_2$ & Bi & Te$_1$ &  \\
\hline
E$_u^1$  & 1.62 &  1318.15 &  -  & 0.310&-0.475 & 0.596& -0.475 & 0.310 & $x$, $y$\\
E$_u^2$   & 2.75 &  15.48 &   -  & 0.494& -0.114& -0.696& -0.114& 0.494& $x$, $y$\\
A$_{2u}^1$ & 2.79  &  - &   540.46 & 0.122& 0.262& -0.913& 0.262& 0.122& $z$\\
A$_{2u}^2$ & 3.60  &  - &  273.73 & 0.571& -0.413& -0.085& -0.413& 0.571& $z$\\
\hline
\end{tabular}
\end{table}

The oscillator strengths $S_\perp$ and $S_\parallel$ for an electric field respectively
perpendicular and parallel to the trigonal axis are reported 
in table \ref{osc_strength_tab}. We have previously seen that the oscillator strengths
are related to both the eigenvectors of the zone center dynamical matrix and the mode effective
charges (See Eq. \ref{def_X}).
The huge difference between the oscillator strength
for the $E_u^1$ and $E_u^2$ modes arises from the fact that all atoms contribute to 
$S_\perp$ for the $E_u^1$ mode while only the Bi atoms contribute significantly 
for the  $E_u^2$ mode. Indeed, the displacements of atoms have the same sign for 
both modes with the exception of the Te$_2$ atom whose displacement along the $E_u^2$ mode
is reversed, producing an almost complete cancellation of the Te atoms to $S_\perp$. 
\begin{figure}[!htbp]
\begin{center}
\vskip0.5truecm
\includegraphics[angle=0, scale=0.6]
{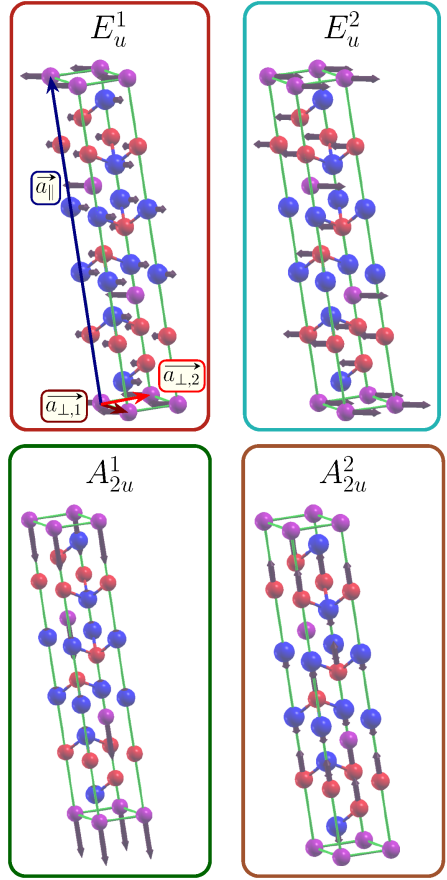}
\end{center}
\caption{
Displacements of the atoms for the two $E_u$ modes polarized along the $x$-axis and
for the two $A_{2u}$ modes polarized along the $z$-axis (trigonal axis).
The Te$_1$, Te$_2$ and Bi atoms are respectively colored in red,
purple and blue.
}
\label{Eigenvectors_fig}
\end{figure}
For the sake of completeness, the displacements of atoms inside the hexagonal 
unit cell for the four different IR active modes are
depicted in Fig. \ref{Eigenvectors_fig}.
\begin{figure}[!htbp]
\begin{center}
\vskip0.5truecm
\includegraphics[angle=0, scale=0.5]
{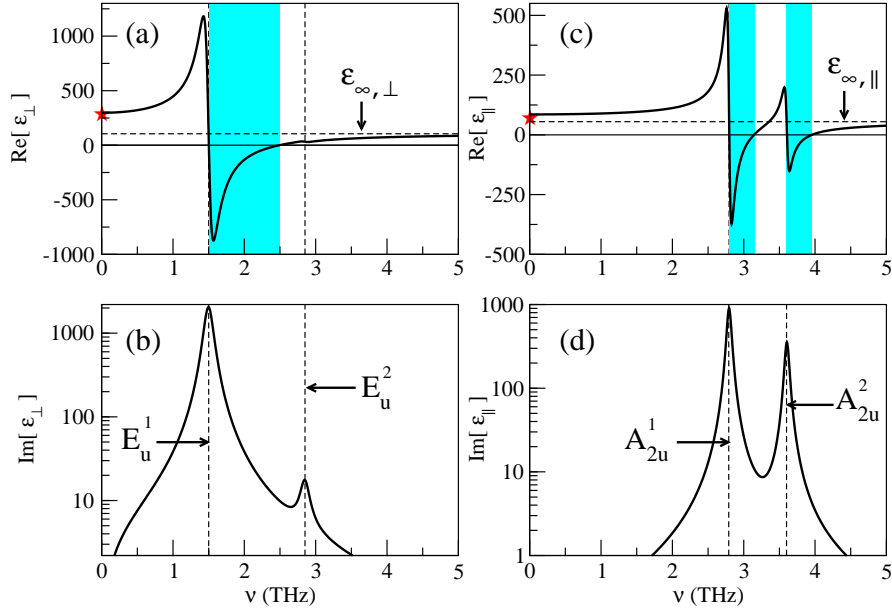}
\end{center}
\caption{Imaginary and real part of the  IR dielectric
function of Bi$_2$Te$_3$ computed with Eq. \ref{chi_ionic_def2} 
for an electric field perpendicular 
to the trigonal axis (pannels a and b) and parallel to the trigonal axis
(pannels c and d) as a function of frequency $\nu$ (in THz). We assumed that 
$\gamma_\lambda=0.44$ THz for the $E_u$ modes and that $\gamma_\lambda=0.22$ THz 
for the $A_{2u}$ modes. Note that the frequencies of both $E_u$ modes have been 
slightly renormalized to match the experimental frequencies\cite{richter_1977} reported in
table \ref{frequency_tab} while the theoretical frequencies of both $A_{2u}$ modes have been
considered. The stars in pannel (a) and (c) denote the experimental static dielectric 
constants extracted from the IR reflectance spectras measured at 15 K\cite{richter_1977}
and the 
colored domain in pannel (a) delimits the Restrahlen band of the $E_{u}^1$ 
mode situated between the transversal optical frequency at 1.5 THz 
and the longitudinal optical frequency at 2.49 THz while the
two colored domains in pannel (c) delimit the Restrahlen bands of the $A_{2u}^1$ 
($A_{2u}^2$) mode situated between the transversal optical frequency at 2.79 THz (3.60 THz)
and the longitudinal optical frequency at 3.14 THz (3.95 THz).
}
\label{Im_Re_IR_epsilon_fig}
\end{figure}

The IR dielectric constant is defined as
\begin{equation}\label{def_epsilon_ionic}
\epsilon_{\perp, \parallel}(\nu)= \epsilon_{\infty, \perp, \parallel} + 
\chi^{ion}_{\perp, \parallel}(\nu),
\end{equation}
where the electronic contribution to the static dielectric function at the RPA level 
denoted as $\epsilon_{\infty, \perp, \parallel}$, also named clamped-nuclei dielectric constant,
can be approximated as a constant for frequencies less than 5 THz. 
Fig. \ref{Im_Re_IR_epsilon_fig}(a) and \ref{Im_Re_IR_epsilon_fig}(c) display the real part
of the IR dielectric function computed for an electric field perpendicular 
(${\bf{E}} \perp {\bf{c}}$) and parallel (${\bf{E}} \parallel {\bf{c}}$) 
to the trigonal axis.
The two $E_u$ modes can be driven for ${\bf{E}} \perp {\bf{c}}$ while the two
$A_u$ modes can be driven for ${\bf{E}} \parallel {\bf{c}}$. The real part changes
sign each time the frequency $\nu$ crosses the frequency of an IR active mode 
depicted as a vertical dashed line in Fig. \ref{Im_Re_IR_epsilon_fig}. 
The only exception concerns the mode $E_u^2$ whose oscillator strength is very
weak (see Table \ref{osc_strength_tab}). Interestingly, our calculations reproduce 
the static dielectric functions denoted as stars 
for both polarizations. The ionic and electronic contributions to the total susceptibility
are respectively $\sim$ 192 and $\sim$ 104  ($\sim$ 29 and $\sim$ 55) 
for ${\bf{E}} \perp {\bf{c}}$ (${\bf{E}} \parallel {\bf{c}}$), reflecting the high degree
of anisotropy of the optical properties in the THz range. It's worth outlining that
$\textrm{Re}[\epsilon_\perp]. \textrm{Re}[\epsilon_\parallel]< 0$ for the three Restrahlen 
bands depicted as vertical colored domains 
in Fig. \ref{Im_Re_IR_epsilon_fig}. 
Hence, Bi$_2$Te$_3$ as h-BN\cite{Caldwell_2014} 
or $\alpha$-MoO$3$\cite{Ma_2018} exhibits natural hyperbolicity 
and might support hyperbolic phonon-polaritons. We also note that the imaginary part of the 
IR dielectric function displayed in Fig. \ref{Im_Re_IR_epsilon_fig}(b) and 
\ref{Im_Re_IR_epsilon_fig}(d) has a lorentzian profile in the vicinity of each resonance 
with a full width at half maximum (FHWM) given by $\gamma_\lambda/\pi$ and remind that
the phonon lifetime is given by $1/\gamma_\lambda$. Here, we assumed that the lifetimes
of both $E_u$ modes is $\sim 2.27$ ps while the lifetimes of both $A_u$ modes is 
$\sim 4.54$ ps to best reproduce the IR reflectance spectras discussed in the following
section.

\section{INFRARED REFLECTIVITY SPECTRUM}\label{IR_reflectivity} 
\begin{figure}[!htbp]
\begin{center}
\vskip1.0truecm
\includegraphics[angle=0, scale=0.4]
{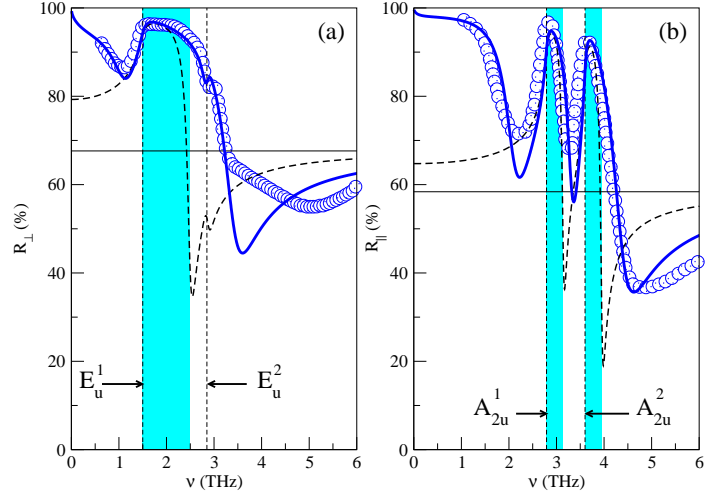}
\end{center}
\caption{
Computed IR reflectivity spectras with (thick lines) and without (dashed lines) the
free electron contribution for ${\bf{E}} \perp {\bf{c}}$ (left pannel) 
and ${\bf{E}} \parallel {\bf{c}}$ (right pannel) compared to the experimental spectras
 (open circles) recorded at 15 K\cite{richter_1977}. 
The restrahlen bands are depicted as colored domains and the horizontal lines correspond
to the expected value of the reflectivity well above the highest IR frequency 
($\sim 3.6$ THz) but below the band gap frequency ($\sim 51$ THz).
}
\label{Reflectivity_IR_fig}
\end{figure}

The normal incidence reflectivity spectras computed according to Eq. \ref{def_R_calc} 
for ${\bf{E}} \perp {\bf{c}}$ and 
${\bf{E}} \parallel {\bf{c}}$ are compared to the experimental spectras 
at 15 K\cite{richter_1977} in Fig. \ref{Reflectivity_IR_fig}.
Roughly speaking, the overall agreement between the calculated reflectivity (dashed lines)
and the experimental reflectivity (open circles) for both polarizations is fairly good 
within the Restrahlen bands, while poor outside, suggesting that the free
electrons arising from doping contribute to the dielectric function.
Hence, we consider the dielectric function
\begin{equation}
\overline{\epsilon}_{\perp, \parallel}(\nu)=\epsilon_{\perp, \parallel}(\nu) 
- \nu_{pl, \perp, \parallel}^2/\nu (\nu + i\gamma_{\perp, \parallel}),
\end{equation}
where $\epsilon_{\perp, \parallel}$ is defined in Eq. \ref{def_epsilon_ionic} and 
where the charge carrier contribution depends on two parameters, namely the plasmon
frequency $\nu_{pl, \perp, \parallel}$ and plasmon damping $\gamma_{\perp, \parallel}$.
The calculated reflectivity shown as a solid line in Fig. \ref{Reflectivity_IR_fig} 
for ${\bf{E}} \perp {\bf{c}}$ (${\bf{E}} \parallel {\bf{c}}$)
with $\nu_{pl, \perp}\sim 26.6$ THz ($\nu_{pl, \parallel}\sim 24.2$ THz)
reproduces nicely the experimental reflectivity provided that the plasmon
damping constant $\gamma_\perp$ ($\gamma_\parallel$) is well chosen. We found that
a constant plasmon damping $\gamma_\perp \sim 0.95$ THz allows to reproduce the 
experimental reflectivity spectra for ${\bf{E}} \perp {\bf{c}}$ while a frequency dependent
plasmon damping $\gamma_\parallel$ is crucial to reproduce the reflectivity spectra for 
${\bf{E}} \parallel {\bf{c}}$.
Hence, we considered the following formula:
\begin{equation}\label{damping_fitting}
\gamma_\parallel(\nu)=\frac{\gamma_{l}+\gamma_{h}}{2} +
\frac{\gamma_{h}-\gamma_{l}}{2} \tanh\left(\frac{\nu -\nu_0}{\Delta \nu}\right),
\end{equation}
to interpolate
between the low frequency value of the plasmon damping denoted as $\gamma_l \sim 0.24$ THz 
and the high frequency value, denoted as $\gamma_h \sim 1.7$ THz with a fairly 
steep variation in a frequency
interval $2 \Delta \nu \sim$ 1.4 THz centered on $\nu_0 \sim 2.3$ THz.

 

\section{PLASMON FREQUENCIES}\label{plasmons} 
As shown in the previous part, the Drude contribution to the IR reflectivity can
not be ignored. Indeed, a bulk charge carrier contribution originates from intrinsic
defects, such as anion vacancies and antisite defects that 
are ubiquitous in most compound semiconductors\cite{Chuang_2018}. 
For instance, focusing on the latter,  
either a Te$_1$ atom can be replaced by a Bi atom, leading to hole charge carriers or
a Bi atom can be replaced by a Te atom, leading to electron charge carriers. From an
experimental point of view, the p-type or n-type charge carrier concentration can 
range from $3\times 10^{17}$ to $5\times 10^{19}$ cm$^{-3}$, depending on the growth 
conditions.
\begin{figure}[!htbp]
\begin{center}
\vskip1.0truecm
\includegraphics[angle=0, scale=0.4]
{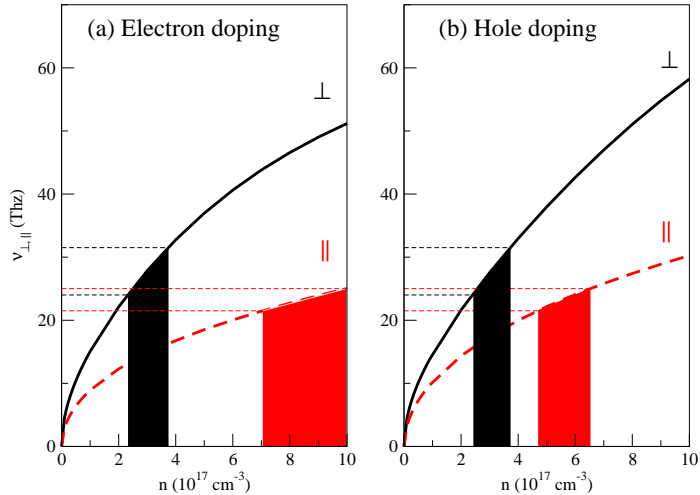}
\end{center}
\caption{
Plasmon frequencies $\nu_{pl, \perp}$ and $\nu_{pl, \parallel}$ (in THz) computed as
a function of electron doping (left pannel) and hole doping (right pannel) up to $n=10^{18}$
cm$^{-3}$.
The horizontal dashed lines of the same color span the range of plasmon frequencies
that gives a good fitting of the reflectivity spectras shown in Fig. \ref{Reflectivity_IR_fig}.
The range of doping corresponding to the range of $\nu_{pl, \perp}$ ($\nu_{pl, \parallel}$)  is
depicted in black (red). The plasmon frequencies, that are hard to converge, are computed
using a very dense $256\times 256 \times 256$ grid of kpoints for a temperature $T\sim 15$ K.
}
\label{Plasmons_fig}
\end{figure}
Assuming that the rigid band model is valid, the plasma frequencies can be evaluated
by tuning the Fermi level to achieve either electron or hole doping. In such a case,
it is straightforward to show that the intraband contribution to the dielectric tensor  
defined in Eq. \ref{macroscopic_tensor} 
behaves as $-\overline{\overline{\omega_{pl}^2}}/\omega^2$ where:
\begin{equation} \label{omega2}
\omega_{pl, \alpha, \beta}^2=\frac{e^2}{4\pi\epsilon_0}\frac{4 \pi}{v} 
\frac{1}{N}\sum_{{\bf k},n}
v_{n{\bf k}}^\alpha v_{n{\bf k}}^\beta 
\left( -\frac{\partial f}{\partial \epsilon}\right)_{\epsilon_{n{\bf k}}}.
\end{equation}
Here ${\bf v}_{n{\bf k}}\equiv \nabla_{\bf k} \epsilon_{n{\bf k}}/\hbar$ 
are the band velocities. It is important to note that only states at
the Fermi level contribute to $\omega_{pl, \alpha, \beta}^2$ at $T=0$ since
$-\frac{\partial f}{\partial \epsilon} \to \delta[\epsilon-\epsilon_F]$ 
when $T \to 0$ and that $\omega_{pl, \alpha, \beta}^2$ is expected
to increase when $T$ increases because of the larger number of states 
contributing to the sum in Eq. \ref{omega2}. Replacing
$\epsilon_{n{\bf k}}$ by $\hbar^2 k^2/2m$ in Eq. \ref{omega2} 
is tantamount to considering a free electron gas and leads
to the classical equation $\omega_{pl}^2=n e^2/m \epsilon_0$ 
where $n$ is the electron density number. In such a case, we obtain
an isotropic plasma frequency. For Bi$_2$Te$_3$, the situation is clearly different.
Indeed, $\overline{\overline{\omega_{pl}^2}}$ is a diagonal tensor with only
two independent elements, namely 
$\omega_{pl, \perp}^2 \equiv \omega_{pl, 1, 1}^2=\omega_{pl, 2, 2}^2$ and
$\omega_{pl, \parallel}^2 \equiv \omega_{pl, 3, 3}^2$. Furthermore, both the
last valence and first conduction band near the six extremas discussed 
in part \ref{electronic_structure} display a non-parabolic dispersion as shown in
Shubnikov-de Haas experiments\cite{kohler1_1976, kohler2_1976}. We computed
the plasmon frequencies 
$\nu_{pl,\perp, \parallel}=\omega_{pl,\perp, \parallel}/2\pi$ as a function
of doping $n$ with $n$ up to $10^{18}$ cm$^{-3}$, corresponding to a low level
of doping ($\sim 1.69 \times 10^{-4}$ electrons or holes per unit cell). As
shown in Fig. \ref{Plasmons_fig}, $\nu_{pl,\perp}$ is larger 
than $\nu_{pl,\parallel}$ for each $n$ and they both increase when $n$ increases.
However, $\nu_{pl,\parallel}$ increases faster for hole doping 
(See Fig. \ref{Plasmons_fig}(b)) than for electron doping 
(See Fig. \ref{Plasmons_fig}(a)) while $\nu_{pl,\perp}$ does not depend on 
the type of doping up to $n_c=4 \times 10^{17}$ cm$^{-3}$ and becomes larger
for hole doping, as compared to electron doping, above $n_c$. 
Interestingly, we reported 
in Fig. \ref{Plasmons_fig} the range of $\nu_{pl,\perp}$ 
(see the dashed black lines) and $\nu_{pl,\parallel}$ (see the dashed red lines)
respectively giving a fairly good agreement between theory and experiment 
for $R_\perp$ and $R_\parallel$. The corresponding values of $n$ are colored 
in black (red) for $\nu_{pl,\perp}$ ($\nu_{pl,\parallel}$). A nice agreement
between our fitted and theoretical values involves an overlap between the two 
colored domains. Neither the calculations for electron doping nor the 
calculations for hole doping produce such an overlap. However, the two colored
domains in Fig. \ref{Plasmons_fig}(b) 
are closest to each other
, suggesting that the sample used
in the experiments\cite{richter_1977} might be hole doped 
with 
$n \sim 4 \times 10^{17}$ cm$^{-3}$. 
While our approach to model doping is
very crude, our calculations suggest that a measure of the reflectivity spectras
in the THz range for both polarizations might be helpful to infer both type and 
level of doping. The validity of our approach might be assessed by cross-checking
the results with Hall measurements.

\section{CONCLUSION}
We studied the optical properties of Bi$_2$Te$_3$ in both the visible and IR range. 
Our first-principles calculations reveal that the dielectric functions computed at the RPA level are
rather anisotropic (a fact disregarded in ellipsometry measurements\cite{dubroka_2017}) 
and strongly impacted by the LF effects when the electric field is polarized along the
trigonal axis. The agreement between our calculated near-normal incidence reflectivity spectra
for an electric field perpendicular to the trigonal axis and 
the experimental spectra\cite{Greenaway_1965} is fairly good, assessing the validity
of our approach to compute the optical spectras by simply shifting the conduction 
bands with respect to the valence bands from 120 meV to roughly mimic the self-energy
corrections. The clamped nuclei static dielectric constants including LF effects
($\epsilon_\perp^\infty\sim 105$ and $\epsilon_\parallel^\infty\sim 56$) are very 
large because of strong direct interband transitions starting from the
direct band gap $\sim 210$ meV and extending up to 4 eV, 
suggesting that excitonic effects can be neglected
in the optical calculations. The ionic contributions to the static dielectric constants,
that can be computed from both the Born effective charges and the IR phonon frequencies
and eigenvectors, are also very large 
as $\chi_\perp^{ion}\sim 192$ and $\chi_\parallel^{ion}\sim 29$. The huge value for 
$\chi_\perp^{ion}$ reflects the strong coupling between an electric field perpendicular
to the trigonal axis and the $E_u^1$ mode. Furthermore, the calculated reflectivity 
spectras in the THz range agree fairly well with the experimental spectras\cite{richter_1977}
provided that the unavoidable contribution of the free carriers arising from defects 
is taken into account by considering a Drude contribution parametrized 
by plasmon frequencies and plasmon damping that are fitted to reproduce the experimental
spectras. We showed that the plasmon frequencies can be computed within the rigid band
approximation as a function of both type and level of doping. Thus, a measurement of
the reflectivity in the THz range for both polarizations might offer the opportunity
to infer not only the type of doping but also the level of doping. Finally, this
work is a first step towards the understanding of the mechanisms governing the
generation of the coherent $A_{1g}$ phonon 
in a THz excited Bi$_2$Te$_3$ nanofilm\cite{Levchuk_2020}
as a non linear coupling between the $E_u^1$ mode driven by the THz pulse 
and the $A_{1g}$ mode might
be at the heart of the experimental observations\cite{Mankowski_2016}.

\begin{acknowledgments}
This work was performed using HPC
resources from GENCI-TGCC, CINES, IDRIS (project AD010905096R1).
\end{acknowledgments}

\end{document}